\documentclass[12pt]{article}
\usepackage{times}
\begin{document}
\vspace*{2cm}
\fontsize{24}{38}\selectfont
\renewcommand{\baselinestretch}{1.27}
\begin{center}
\huge Quantum reference systems: reconciling locality with quantum
mechanics
\end{center}
\renewcommand{\baselinestretch}{1}
\fontsize{12}{24}\selectfont
\vspace{0.9mm}
\begin{center}
Gyula BENE\\
\it Institute for Theoretical Physics, E\"otv\"os University,
P\'azm\'any P\'eter s\'et\'any 1/A, H-1117 Budapest, Hungary
\end{center}
\begin{center}
\parbox[t]{12cm}{\renewcommand{\baselinestretch}{1}\fontsize{10}{22}\selectfont \bf \noindent
  Abstract. \rm The status of locality in quantum mechanics is
  analyzed from a nonstandard point of view. 
It is assumed that quantum states are relative in the
  sense that they 
  depend on and are defined with respect to
  some bigger physical system which contains the former system 
  as a subsystem. Hence, the bigger system acts as a reference
  system. It is shown that quantum mechanics can be
  reformulated in accordance with this new physical assumption. 

  Additional
  laws express the (probabilistic) relation among
states which refer to different quantum reference systems. They
  replace von Neumann's postulate about the measurement (collapse of
  the wave function). The
  dependence of the quantum states on the quantum reference systems 
 resolves the apparent contradiction connected with the measurement
  (Schr\"odinger's cat paradox). There is another important consequence
of this dependence: states may not be comparable, i.e., they cannot
be checked by suitable measurements simultaneously. This special
  circumstance is fully reflected mathematically by the theory. 
Especially, it is shown that certain joint probabilities (or the corresponding combined
  events) which play a vital role in any proof of Bell's theorem
    do not exist. The conclusion is that the principle of locality holds true in
  quantum mechanics,
and one has to give up instead of locality an intuitively  
natural-looking feature of
  realism, namely, the comparability of existing states.
}
\end{center}
\renewcommand{\baselinestretch}{1}
\fontsize{12}{24}\selectfont
\newpage
\section{Introduction}
During the development of physics research is extended 
to phenomena at such spatial, time or energy scales that are  
far beyond the range of everyday experience.
These phenomena sometimes force us to revise our previous concepts which have seemed to us
natural or even indispensable, but are actually rooted in our limited 
previous experience and finally prove to be of approximate
validity. This is reflected in the growing level of abstraction of physical theories.   
Quantum phenomena have forced already several such revisions. A
primary example is the surrender of determinism, but
the wavelike behaviour of particles, the existence of discrete energy levels and Heisenberg's
uncertainty relations also imply substantial revisions of some basic
features of classical physics. It is well known that quantum phenomena 
force us to surrender or revise at least one more very basic concept: Bell's
theorem tells us that that we have to give up (or revise) either locality, or
realism, or inductive inference\cite{hid}-\cite{Bell5}. On the other
hand, these
concepts are so deeply rooted in scientific thinking that one is
reluctant to give up any of them. Many people think that perhaps
locality is the weakest concept of the three (cf.
arguments for nonlocality in Ref.\cite{Stapp} and counterarguments  
in Ref.\cite{Merminrep}). Nevertheless, all the
fundamental equations of physics satisfy the principle of locality,
including e.g. the standard model of the elementary particles, so it is
rather implausible that locality would be violated just in quantum
measurements. The myth of nonlocality is much less attractive today 
than it has been in the fifties or sixties when people tried to build
nonlocal quantum field theories. 

There are at least two other myths related to quantum mechanics: one of them is 
the notorious belief that around some large particle number or mass
quantum mechanics gradually becomes invalid and classical mechanics
starts to be correct. Actually this very simple view has no
experimental support at all. On the contrary, all the available
experimental results confirm the validity of quantum mechanics, even in the 
macroscopic situation of superfluid He. The actual motivation
of the above view is the paradox of quantum measurements, which
involves the third myth, the fictitious collapse  
of the wave function\cite{von Neumann}.
This process was invented for the interpretation of quantum mechanics,
i.e., for making contact between theoretical results and the
experience. Nevertheless, if considered an actual physical process,
it contradicts the Schr\"odinger equation and violates the principle
of locality. 

In the present paper we show that a consistent theory can be
constructed without accepting any of the three myths above. 
As a guiding principle, we try to confine the number of assumptions to a minimum 
and base our considerations on logical
clarity and consistency. Obviously, this is all we can safely do
in a range of phenomena where our intuition does not work. 
As
expected, one has to revise rather substantially some basic concepts.
Namely, it will be shown that one has to give up the belief that
a physical system at a given instant of time has a quantum
state in an absolute sense, i.e., a state which depends only on the 
system to be described    
(apart from unitary transformations which may appear
if another coordinate system is chosen). Instead, quantum states 
will express basicly a relation between a system to be described 
and another system containing the former one. The latter, bigger  
system acts as reference system\cite{Bene}. This new kind of
dependence on reference systems will be explained 
in detail in Section 2. The idea that quantum states are relational  
has been put forward (in mathematically and conceptually different ways) 
in Refs.\cite{Everett},\cite{Rovelli}. The relational nature of 
quantum mechanics is expressed also by the proposal that correlations 
(rather than states) are the basic entities ("correlations 
without correlata") \cite{Mermin}. 
 
In Section 3. the basic rules of quantum mechanics are formulated 
in a way that takes into account the dependence of the states on 
quantum reference systems. von Neumann's measurement postulate 
will be replaced by the rules which express the indeterministic  
relation among states defined with respect to different quantum reference 
systems. The most important of these rules, which has no counterpart 
in standard quantum mechanics, utilizes the eigenstates of the reduced 
density matrix. These eigenstates (Schmidt states\cite{Schmidt}) 
play a central role also in the modal interpretations\footnote{One may even consider 
the present approach a modal interpretation in the sense that it also 
aims at giving a physical meaning to the quantum formalism ("properties attribution") 
like modal interpretations.}\cite{modal}. 
Their significance  
has also been emphasized within the framework of decoherence theory 
\cite{Zeh}, \cite{Albrecht}. 
According to the present approach, Schr\"odinger's equation is universally 
valid and measurements are considered as any other interaction 
between two physical systems. If the dynamics of this interaction 
has the special form of a perfect (von Neumann type) measurement, 
the usual quantum mechanical expressions for the probabilities 
are recovered.  

The dependence of quantum states on quantum reference systems (as defined 
in the present paper) leads to an
even more radical consequence: some states which exist may not be compared.
This is totally unusual for our intuition and classical experience,
but if the states which we would like to compare are defined with
respect to different reference systems, there is no logical necessity
that one can indeed compare them. This basically influences the
picture we have about reality. We cannot think of existing states 
as of letters in a big book. If two letters exist, we may compare them
in principle, without changing them or their relationship. Thus (if we
do not possess complete knowledge about them) it
makes sense to speak about the joint probability of their simultaneous
presence. In case of existing quantum states one may check any of them
by a suitable measurement, without disturbing that particular state.  
But such a measurement usually does disturb the other states, 
therefore, these measurements usually cannot be
performed simultaneously (or one by one)  
without influencing any of the relevant states.
Thus one cannot give any operational definition of the joint
probability of the simultaneous presence of states. This seems to be
rather odd, and one might think that independently of this situation
such probabilities must exist on the ground that each existing state
somehow corresponds to an element of the reality. On the other hand,
one should keep in mind that this set-like picture of the reality
is based on the fact that in classical
physics the property of comparability is always given.
In the absence of that one has no reason to expect that
the consequences remain intact.
Let us emphasize that this modification of the concept of reality
is completely independent of any influence of
consciousness. Therefore, the resulting reality concept 
still preserves the idea that the world is made up of objects
whose existence is independent of human consciousness.
The non-comparability of states will be discussed in Section 4. 
 
In Section 5. the derivation of Bell's inequality is analyzed 
from the point of view of the present modified quantum mechanics. 
It will be shown that the role of the "hidden variable" 
is played by a particular quantum state. 
It turns out that the usual "shadowing property" which 
would express local realism, does not follow, because 
the corresponding three-fold joint probability is physically meaningless 
due to the non-comparability of the relevant quantum states.  
On the other hand, all reasonable 
requirements of locality are fulfilled. 
 
In Section 6. we summarize the results and conclude.

\section{The basic new concept: quantum reference systems}

Let us consider a simple example, namely, an idealized measurement
of an $\hat S_z$ spin component of a spin-$\frac{1}{2}$ 
particle. Be the particle $P$ initially in the state
\begin{eqnarray}
\alpha |\uparrow>+\beta|\downarrow>\;,\label{u1}
\end{eqnarray}
where $|\alpha|^2+|\beta|^2=1$ and the states $|\uparrow>$ and $|\downarrow>$ are the
eigenstates of $\hat S_z$ corresponding to the eigenvalues
$\frac{\hbar}{2}$ and $-\frac{\hbar}{2}$, respectively.
The other quantum numbers and variables have been suppressed. The dynamics of the
measurement is given by the relations
$|\uparrow>|m_0>\; \rightarrow \; |\uparrow>|m_{\uparrow}>$ and  
$|\downarrow>|m_0>\; \rightarrow \; |\downarrow>
|m_{\downarrow}>$, where $|m_0>$ stands for the state 
of the measuring device $M$ (e.g. a Stern-Gerlach apparatus) 
before the
measurement (no spot on the photographic plate),
while $|m_{\uparrow}> $ ($|m_{\downarrow}>$) 
is the state of the measuring device after the measurement
that corresponds
to the measured spin value $\frac{\hbar}{2}$ ($-\frac{\hbar}{2}$).
The shorthand notation $\rightarrow$ stands 
for the unitary time evolution during
the measurement, which is assumed to 
fulfill the time dependent Schr\"odinger equation 
corresponding to the total Hamiltonian of the combined 
$P+M$ system. As the initial state of the particle is
given by Eq.(\ref{u1}), the linearity of the Schr\"odinger equation
implies that the measurement process can be 
written as
\begin{eqnarray}
(\alpha |\uparrow>+\beta |\downarrow>)|m_0> 
\;\rightarrow \; |\Psi>=\alpha |\uparrow>|m_{\uparrow}>
+\beta |\downarrow>|m_{\downarrow}>\quad.
 \label{u2}
\end{eqnarray} 
This simplified dynamics is called a von Neumann type perfect measurement.
Let us consider now the state of the measuring device $M$ after the measurement. As the
combined system $P+M$ is in an entangled state, the 
measuring device has no own wave function and may be described
by the {\em reduced density matrix}\cite{Landau} 
\begin{eqnarray}
\hat \rho_M=Tr_P\left(|\Psi><\Psi|\right)
=|m_{\uparrow}>|\alpha|^2<m_{\uparrow}|
+ |m_{\downarrow}>|\beta|^2<m_{\downarrow}|\quad,\label{u3}
\end{eqnarray}
where $Tr_P$ stands for the trace operation in the Hilbert
space of the particle $P$. Nevertheless, if we look at the measuring device,
we certainly see that either $\frac{\hbar}{2}$ or $-\frac{\hbar}{2}$ spin component
has been measured, that correspond to the states $|m_{\uparrow}> $ and $|m_{\downarrow}>$,
respectively. These are obviously not the same as the state (\ref{u3}).
Indeed, $|m_{\uparrow}> $ and $|m_{\downarrow}>$ are pure states
while $\hat \rho_M$ is not.  
As is well known by now, $\hat \rho_M$ cannot be considered 
a statistical mixture of $|m_{\uparrow}> $ and $|m_{\downarrow}>$, 
i.e., "ignorance interpretation" cannot solve 
the problem of objectification\cite{Mittelstaedt}. 
Why do we get different states?
According to orthodox quantum mechanics,
one may argue as follows. The reduced density matrix $\hat \rho_M$ 
has been calculated from the state $|\Psi>$ (cf. Eq.(\ref{u2}))
of the whole system $P+M$. A state is a result of a measurement
(the preparation), so we may describe $M$ by $\hat \rho_M$ if we have
gained our information about $M$ from a measurement done on $P+M$. 
On the other hand, looking at the measuring device directly
is equivalent with a measurement done directly on $M$.
In this case $M$ is described by either $|m_{\uparrow}> $ or $|m_{\downarrow}>$.
We may conclude that performing  measurements on 
different systems (each containing the system we want to decribe)
gives rise to different descriptions 
(in terms of different states).
Let us call the system which has been measured (it is $P+M$
in the first case and $M$ in the second case) the {\em quantum
reference system}. Using this terminology, we may tell that
we make a measurement on the quantum reference system $R$, thus we prepare
its state $|\psi_R>$ and using this information we calculate
the state $\hat \rho_S(R)=Tr_{R\setminus S} |\psi_R><\psi_R|$ 
of a subsystem $S$. We shall call
$\hat \rho_S(R)$  the state of $S$ with respect to $R$. 
Obviously $\hat \rho_R(R)=|\psi_R><\psi_R|$, thus $|\psi_R>$
may be identified with the state of the system $R$ 
with respect to itself.

Let us emphasize that up to now, despite of the new terminology,
there is nothing new in the discussion. We have merely 
considered some rather elementary consequences of basic quantum mechanics. 

Let us return now to the question why the state of
the system $S$ (i.e., $\hat \rho_S(R)$) depends
 on the choice of the quantum reference
system $R$. In the spirit of the Copenhagen interpretation
one would answer that in quantum mechanics measurements
unavoidably disturb the systems, therefore, if we perform
measurements on  different surroundings $R$, this disturbance is
different, and this is reflected in the $R$-dependence of $\hat \rho_S(R)$.
Nevertheless, this argument is not compelling. We may also
take the realist point of view and  
assume that the states of the systems have already existed before the
measurements, and appropriate measurements do not change
these states.  Then the $R$-dependence
of $\hat \rho_S(R)$ becomes an inherent property of quantum mechanics, 
the states themselves represent actually existing properties (or correspond 
to some elements of the reality), and quantum formalism becomes a desciption 
of the reality rather than a calculational tool which relates 
consecutive measurement results.
Let us leave at this decisive point the traditional framework of quantum mechanics
and follow the new line just sketched. Below a comparison between 
the two approaches is given.  
 \noindent
\begin{tabular}{|p{0pt}p{6.7cm}p{0pt}|p{0pt} p{6.7cm}|}
\hline
&{\bf Standard QM:} &&& {\bf Present approach:} \\\hline
&Quantum reference system dependence is caused by the influence 
of the measurement. Measurement is a primary concept. &&&
 
Quantum reference system dependence is a fundamental property. 
States exist and depend on quantum reference systems even in the 
absence of any measurement. Measurement is a derived concept.\\ 
\hline
\end{tabular}
\vskip0.5cm
 
The meaning of the quantum reference systems is now analogous
to the classical coordinate systems. Choosing a
classical coordinate system means that we imagine
what we would experience if we were there. Similarly,
choosing a quantum reference system $R$ means that we
imagine what we would experience if we performed a measurement on
$R$ that does not disturb $\hat \rho_R(R)=|\psi_R><\psi_R|$. 
In order to see that such a measurement exists, consider an
operator $\hat A$ (which acts on the Hilbert space of $R$) whose
eigenstates include $|\psi_R>$. The measurement of $\hat A$
will not disturb $|\psi_R>$. Let us emphasize that the
possibility of nondisturbing measurements is an expression
of realism: the state $\hat \rho_R(R)$ exists independently
whether we measure it or not.  
 
Despite the analogy, there are several differences between 
the concept of classical and quantum reference systems. 
These are summarized below: 
 \vskip0.5cm
\noindent
\begin{tabular}{|p{0pt}p{6.7cm}p{0pt}|p{0pt} p{6.7cm}|}
\hline
&{\bf Classical reference system (CRS):}&&& 
 
{\bf Quantum reference system (QRS):}\\ 
 \hline
&CRS is an abstraction of actual physical systems, 
as their detailed structure is not important.&&& 
 
QRS is a physical system. The detailed structure cannot be 
eliminated. \\
 \hline
&measurements are done on a system by devices attached to the CRS&&& 
 
measurement is done on the QRS \\
 \hline
&there is a one-to-one relationship (transformation) 
between descriptions with respect to two different CRS-s.&&& 
 
there is a stochastic relationship between descriptions 
with respect to two different QRS-s (indeterminism), 
or they may not even be compared \\
 \hline
\end{tabular}
\vskip0.5cm

As the dependence of $\hat \rho_S(R)$ on $R$
is a fundamental property now, one has to specify 
the relation of the different states and has to relate the 
formalism with experience. This can be done in  terms of suitable
postulates\cite{Bene}. Below these postulates are listed and 
are applied to the theoretical description of measurements.

\section{Rules of the new framework}

{\bf Postulate 1. \em The system S to be described is a subsystem of the
reference system R.} 

{\bf Postulate 2. \em The state $\hat \rho_S(R)$ is a positive definite, 
Hermitian
operator with unit trace, acting on the Hilbert space of $S$.}

{\bf Definition 1. \em $\hat \rho_S(S)$ is called the internal state 
of $S$.}

{\bf Postulate 3. \em The internal states $\hat \rho_S(S)$ are always 
projectors, i.e., $\hat \rho_S(S)=|\psi_S><\psi_S|$.}

In the following these projectors will be identified with the 
corresponding wave functions $|\psi_S>$ (as they are uniquely
related, apart from a phase factor).
 
{\bf Postulate 4. \em The state of a system $S$ with respect to the
reference system $R$ (denoted by $\hat \rho_S(R)$) is the reduced
density matrix of $S$ calculated from the internal state of $R$,
i.e.
$\hat \rho_S(R)=Tr_{R\setminus S} \left(\hat \rho_R(R)
\right)$,
where $R\setminus S$ stands for the subsystem of $R$ 
complementer to $S$.}

{\bf Definition 2.\em  An isolated system is 
such a system that has not been interacting 
with the outside world. A closed system
is such a system that is not interacting with any other
system at the given instant of time 
(but might have interacted in the past).}

{\bf Postulate 5.\em  If $I$ is an isolated system then its state is 
independent of the reference system $R$:}
$
\hat \rho_I(R)=\hat \rho_I(I)$.

{\bf Postulate 6.\em  If the reference system $R=I$ is an isolated one 
then the state $\hat \rho_S(I)$ commutes with the
internal state $\hat \rho_S(S)$.}

This means that the internal state of $S$ coincides with
one of the eigenstates of $\hat \rho_S(I)$.

{\bf Definition 3.\em  The possible internal states are the eigenstates 
of
$\hat \rho_S(I)$ provided that the reference system $I$ is an
 isolated one.}
 
{\bf Postulate 7. \em If $I$ is an isolated system, then the
probability $P(S,j)$ that the
eigenstate $|\phi_{S,j}>$ 
of $\hat \rho_S(I)$ coincides with $\hat \rho_S(S)$ 
is given by the corresponding eigenvalue $\lambda_j$.} 

{\bf Postulate 8.\em  The result of a measurement is contained 
unambigously in the internal state of the measuring device.}

{\bf Postulate 9. \em If there are $n$ ($n=2,\;3,\;...$) disjointed physical systems, 
denoted by
\hfill\break
$S_1, S_2, ... S_n$, all contained in the isolated reference 
system $I$ and 
having the 
possible internal states
$|\phi_{S_1,j}>, |\phi_{S_2,j}>,...,|\phi_{S_n,j}>$, respectively, 
then the joint
probability that $|\phi_{S_i,j_i}>$ 
coincides with the internal state of $S_i$ ($i=1,..n$)
is given by
\begin{eqnarray}
P(S_1,j_1,S_2,j_2,...,S_n,j_n)\mbox{\hspace{5cm}}\nonumber\\
=Tr_{S_1+S_2+...+S_n} [\hat \pi_{S_1,j_1} 
\hat \pi_{S_2,j_2}
...\hat \pi_{S_n,j_n}\hat \rho_{S_1+S_2+...+S_n}(I)], \label{u5}
\end{eqnarray}
where $\hat \pi_{S_i,j_i}=|\phi_{S_i,j_i}><\phi_{S_i,j_i}|$.}

{\bf Postulate 10.\em  The internal state $|\psi_C>$ of a closed system 
$C$
satisfies the time dependent Schr\"odinger equation}
$i\hbar \partial_t |\psi_C>=\hat H |\psi_C>$.

Here $\hat H$ stands for the Hamiltonian.
\vskip0.5cm
 
Let us emphasize that {\bf Postulate 6} and {\bf 8} do not have any  
counterpart in standard quantum mechanics. They replace the measurement 
postulate but are not simple translations of that (as it will be 
shown below). The relation to the experience 
(that provides the whole construction with a physical meaning)  
is expressed by {\bf Postulate 8}. On the other hand, {\bf Postulate 6} 
and {\bf Postulate 8} are mathematically equivalent with the proposals  
in Refs.\cite{Zeh},\cite{modal}. The present approach differs from these 
at the level of interpretation.
 
In order to demonstrate the working of the postulates  
let us consider again the measurement discussed in the previous 
section. A measurement is treated as a usual interaction 
between two systems, therefore, it is specified by a Hamiltonian 
or the corresponding unitary time evolution. For simplicity we assume 
that it is given by Eq.(\ref{u2}) and that the measuring device  
+ measured object composite is an isolated system. Then {\bf Postulate 5} 
and  {\bf Postulate 3} imply that $\hat \rho_{P+M}(P+M)=|\Psi><\Psi|$. 
According to {\bf Postulate 4} the state of the measuring device 
with respect to the compound system $P+M$ is  
\begin{eqnarray} 
\hat \rho_M(P+M)= 
|m_{\uparrow}>|\alpha|^2<m_{\uparrow}| 
+ |m_{\downarrow}>|\beta|^2<m_{\downarrow}|\quad.\label{u3v} 
\end{eqnarray}  
Applying {\bf Postulate 6} we get that the internal state $\hat \rho_M(M)$ (cf. {\bf 
Definition 1}) of the measuring device is either $|m_{\uparrow}><m_{\uparrow}|$ 
(with probability $|\alpha|^2$, according to {\bf Postulate 7}) 
or $|m_{\downarrow}><m_{\downarrow}|$ (with probability $|\beta|^2$).  
Finally, {\bf Postulate 8} tells us that the actual measurement results (the 
experience) correspond to this $\hat \rho_M(M)$. As one can see, 
we get the same result as in orthodox quantum mechanics   
(this time without assuming the collapse of the wave function). 
One reason for this was the special dynamics (\ref{u2}).  
In orthodox quantum mechanics one usually 
specifies only which observable has been measured. It tacitly assumes 
a simple approximation for the dynamics of the measurement like (\ref{u2}).  
In such cases our postulates lead to the same results.  
If the dynamics is different (e.g., a more detailed description  
is given), nothing ensures 
that one gets back the results of the traditional approach exactly. 
This means that the above postulates cannot be considered as mere 
translations of orthodox quantum mechanics.  
One should also be aware that the present approach leaves much less 
flexibility than orthodox quantum mechanics. While in the traditional 
approach one may rather freely choose a bordering line 
between quantum and classical regime, according to the present 
approach such bordering line does not exist, and one should in principle 
always apply a quantum description. Based on this description 
one may justify the approximations which lead to the traditional approach.      

\section{Non-comparability of the states} 

It is an important feature of the present approach that the states defined
with respect to different quantum reference systems are not
necessarily comparable. Below we explain what it means. 

The present approach works as follows. We assume that the internal
state of an isolated system $I$ is known. Then the states of all the
subsystems $S_j$ of $I$ with respect to $I$ (i.e., $\hat
\rho_{S_j}(I)$) are also known together with their possible internal
states (i.e., the eigenstates of $\hat \rho_{S_j}(I)$). The aim of a
measurement performed on $S_j$ is to learn which of the possible
internal states is the actual one. This can be done without
disturbing that state.
It can be achieved if an operator commuting with $\hat \rho_{S_j}(I)$ is
measured, i.e., if 
the dynamics of the measurement is approximately given by
\begin{eqnarray}
|\phi_{S_j,k}>|m_0>\rightarrow |\phi_{S_j,k}>|m_k>\quad,\label{ndm1}
\end{eqnarray}
where $|\phi_{S_j,k}>$ stands for the $k$-th possible internal state of $S_j$
(cf. {\bf Definition 3}).

If two subsystems $S_1$ and $S_2$ are disjoint ($S_1\cap S_2=
\emptyset$) then one can perform such nondisturbing measurements on
both systems simultaneously without disturbing either  
$\hat \rho_{S_1}(I)$ or $\hat \rho_{S_2}(I)$. Correspondingly, {\bf
  Postulate 9} provides us with a positive definite expresssion
for the joint probability that these states coincide with specified
possible internal states.

There is a special situation if $S_2=I\setminus S_1$. The internal
state of the isolated system can be expressed in this case by the possible
internal states of $S_1$ and $S_2$ as
\begin{eqnarray}
|\psi_I>=\sum_k c_k |\phi_{S_1,k}>|\phi_{S_2,k}>\;.\label{schm}
\end{eqnarray}
This is the Schmidt representation\cite{Schmidt}. It readily implies
that there is a unique relationship between the internal state of
$S_1$ and that of $S_2$. Indeed, Eq.(\ref{u5}) yields
\begin{eqnarray}
P(S_1,j_1,S_2,j_2)=P(S_1,j_1)\delta_{j_1j_2}
\end{eqnarray}
or
\begin{eqnarray}
P(S_2,j_2|S_1,j_1)=\delta_{j_1j_2}\;.\label{condprob}
\end{eqnarray} 
We can rewrite Eq.(\ref{u5}) in the general case 
as follows:
\begin{eqnarray}
P(S_1,j_1, ...S_n,j_n)=<\psi_I|\phi_{S_1,j_1}><\phi_{S_1,j_1}|...|\phi_{S_n,j_n}><\phi_{S_1,j_1}|\psi_I>\;.
\end{eqnarray}
Eq.(\ref{schm}) implies
\begin{eqnarray}
<\psi_I|\phi_{S_1,j_1}><\phi_{S_1,j_1}|=\sum_k  c_k^*
<\phi_{I\setminus S_1,k}|<\phi_{S_1,k}|\phi_{S_1,j_1}><\phi_{S_1,j_1}|\\
= c_{j_1}^* <\phi_{S_1,j_1}|<\phi_{I\setminus S_1,k}|
=<\psi_I|\phi_{I\setminus S_1,j_1}><\phi_{I\setminus S_1,j_1}|
\end{eqnarray}
i.e., one can replace $\hat
\pi_{S_1,j_1}=|\phi_{S_1,j_1}><\phi_{S_1,j_1}|$
by $\hat
\pi_{I\setminus S_1,j_1}=|\phi_{I\setminus S_1,j_1}><\phi_{I\setminus
  S_1,j_1}|$
in Eq.(\ref{u5}). This sometimes makes possible simultaneous nondisturbing
check of internal states of non-disjoint systems.
 
Suppose, e.g., that $I=S_1+S_2+S_3$ (where $S_i\ne\emptyset$ and $S_i\cap S_j=\emptyset$ if $i\ne j$)  
and we would like to learn  
the internal state of $A=S_1+S_2$ and of $B=S_2+S_3$. 
The systems $A$ and $B$ are not disjoint, but their 
complementary systems $S_3$ and $S_1$ are. Therefore, one 
may perform nondisturbing measurements on $S_3$ and $S_1$. 
The knowledge of the internal states of these systems uniquely 
specifies the  
internal states of the original systems $A$ and $B$  
(cf. Eq.(\ref{condprob})). 
Correspondingly, Eq.(\ref{u5}) yields a positive definite 
expression for $P(A,j,B,k)$. 
 
There are, however, situations when neither the systems nor their 
complementary systems are disjoint. The simplest case is 
if $I=S_1+S_2+S_3+S_4$ (where $S_i\ne\emptyset$ and $S_i\cap S_j=\emptyset$ if $i\ne j$)  
and we would like to learn  
the internal state of $A=S_1+S_2$ and of $B=S_2+S_3$. 
Obviously, 
\begin{eqnarray} 
A\ne B\;,A\cap B=S_2\ne \emptyset\\ 
(I\setminus A)\cap B=(S_3+S_4)\cap(S_2+S_3)=S_3\ne \emptyset\\ 
A\cap (I\setminus B)=(S_1+S_2)\cap(S_1+S_4)=S_1\ne \emptyset\\ 
(I\setminus A)\cap (I\setminus B)=(S_3+S_4)\cap(S_1+S_4)=S_4\ne \emptyset 
\end{eqnarray}
Thus, one cannot relate this situation to the case of disjoint subsystems. 
Indeed, a measurement done on $A$ or on $I\setminus A$ (which does not disturb 
$\hat \rho_A(A)$) usually does disturb  $\hat \rho_B(B)$ and  
$\hat \rho_{I\setminus B}(I\setminus B)$. 
Therefore, there is no way to perform a measurement which  
 conveys information about both the internal state of $A$  
 and that of $B$. This is reflected mathematically by the postulates,  
 namely, the positivity of $P(A,i,B,k)$ (if Eq.(\ref{u5}) is 
formally applied) is not guaranteed (it is usually not even real). 
In such a situation the states $\hat \rho_A(A)$, $\hat \rho_B(B)$ 
cannot be compared, although both exist separately. 
Conclusions based on assumptions about the simultaneous 
existence of $\hat \rho_A(A)$ and $\hat \rho_B(B)$ (in the sense 
that one imagines that he knows both states) or, especially,  about 
a related joint probability can lead to contradictions. Therefore, 
non-comparability of existing states is an essential feature 
of the present approach.

\section{Bell's inequality and the principle of locality}

Let us consider a Bell-type experiment performed on  
two spin-$1/2$ particles. 
In order to exhibit
 the mathematical structure we write the internal state of the two particle 
 system before the measurements as
\begin{eqnarray}
\sum_j c_j
|\phi_{P_1,j}>|\phi_{P_2,j}>\label{u14}
\end{eqnarray}
where $c_1=a$, $c_2=-b$, $
|\phi_{P_1,1}>=|1,\uparrow>$, $
|\phi_{P_1,2}>=|1,\downarrow>$, $
|\phi_{P_2,1}>=|2,\downarrow>$, $
|\phi_{P_2,2}>=|2,\uparrow>$.
Eq.(\ref{u14}) is just the Schmidt representation, 
thus
$
P(P_1,j,P_2,k)=|c_j|^2\delta_{j,k}
$.

Let us consider now a typical experimental situation, 
when measurements on both
particles are performed. We shall show
that according to the present theory the observed
correlations are exclusively due to the previous interaction
between the particles. Before the measurements the internal
state of the isolated system $P_1+M_1+P_2+M_2$ ($P_1,P_2$
stands for the particles and $M_1,M_2$ for the measuring
devices, respectively) is given by\hfill\break
$
\left(\sum_j c_j
|\phi_{P_1,j}>|\phi_{P_2,j}>\right)|m^{(1)}_0> |m^{(2)}_0>$,
while it is
\begin{eqnarray}
\sum_j c_j
\hat U_t(P_1+M_1)\left(|\phi_{P_1,j}>|m^{(1)}_0>\right)
\hat U_t(P_2+M_2)\left(|\phi_{P_2,j}>|m^{(2)}_0>\right) \quad,
\label{u15}
\end{eqnarray}
a time $t$ later, i.e. during and after the measurements. Here 
$\hat U_t(P_i+M_i)$ ($i=1,2$) stands for the unitary time evolution operator
of the closed system $P_i+M_i$.

Eq.(\ref{u15}) implies that the internal states of 
the closed systems $P_1+M_1$ and $P_2+M_2$
evolve unitarily and do not influence each
other. This follows readily if one applies 
{\bf Postulates 4, 6} to Eq.(\ref{u15}) and takes into account 
the unitarity of $\hat U_t(P_i+M_i)$.  
This time evolution can be given explicitly through the 
relations
\begin{eqnarray}
|\xi(P_i,j)>|m^{(i)}_0>\;\rightarrow \;|\xi(P_i,j)>|m^{(i)}_j>\quad,
\label{u16}
\end{eqnarray}
where $i,j=1,2$ and $|\xi(P_i,j)>$ is the $j$-th eigenstate of the 
spin measured 
on the $i$-th particle along an axis $z^{(i)}$ which closes an angle
$\vartheta_i$ with the original $z$ direction. The time evolution of 
the internal state of the closed systems $P_i+M_i$ is given explicitly by
$
|\psi_{P_i}>|m_0^{(i)}>\;\rightarrow \;
\sum_j <\phi_{P_i,j}|\psi_{P_i}>|\phi_{P_i,j}>|m_j^{(i)}>
$.
As we see, the $i$-th measurement process 
is completely determined by the initial internal states of the
particle $P_i$. Therefore, any correlation between
the measurements may only stem from the initial correlation
of the internal states of the particles.

For the calculation of the state $\hat \rho_{M_1}(M_1)$ (which corresponds to the
measured value, cf. {\bf Postulate 8}) one needs to know the state of the whole isolated system
$P_1+P_2+M_1+M_2$.
Using Eq.(\ref{u16}) the final state (\ref{u15}) may be written as
\begin{eqnarray}
\sum_{j,k}\left(
\sum_l c_l<\xi(P_1,j)|\phi_{P_1,l}><\xi(P_2,k)|\phi_{P_2,l}>
\right)\mbox{\hspace{4cm}}\nonumber\\
\times |m^{(1)}_j>|m^{(2)}_k>|\xi(P_1,j)>|\xi(P_2,k)>.\nonumber
\end{eqnarray}

Direct calculation shows that
\begin{eqnarray}
\hat \rho_{M_1}(P_1+P_2+M_1+M_2)\mbox{\hspace{5cm}}\nonumber\\
=\sum_j\left(\sum_l
|c_l|^2 |<\xi(P_1,j)|\phi_{P_1,l}>|^2\right)|m^{(1)}_j><m^{(1)}_j|
.\nonumber
\end{eqnarray}
Note that it is independent of the second measurement.

According to {\bf Postulate 6}  $|\psi_{M_1}>$ is one of the
$|m^{(1)}_j>$-s. (Similarly one may derive that
$|\psi_{M_2}>$ is one of the
$|m^{(2)}_k>$-s.)
The probability of the observation of the $j$-th result (up or
down spin in a chosen direction) is
\begin{eqnarray}
P(M_1,j)=\sum_l |c_l|^2 |<\xi(P_1,j)|\phi_{P_1,l}>|^2\;.\label{u17}
\end{eqnarray}
This may be interpreted in conventional terms: $|c_l|^2$ is
the probability that $|\psi_{P_1}>=|\phi_{P_1,l}>$,
and $|<\xi(P_1,j)|\phi_{P_1,l}>|^2$ is the conditional
probability that one gets the $j$-th result if $|\psi_{P_1}>=|\phi_{P_1,l}>$.
Indeed, the initial internal state of $P_1$ uniquely determines 
the internal state of $P_1+M_1$ (owing to the unitary time evolution), 
and the internal state of $P_1+M_1$ uniquely determines 
the internal state of its complementary system $P_2+M_2$ (and vica versa). 
Therefore, the joint probability that the initial  
internal state of $P_1$  
coincides with its $l$-th possible internal state and   
the final internal state of $M_1$  
coincides with its $j$-th possible internal state 
is the same as the joint probability $P(P_2+_2,l,M_1,j)$ that the final  
internal state of $P_2+M_2$  
coincides with its $l$-th possible internal state and   
the final internal state of $M_1$  
coincides with its $j$-th possible internal state. As  
 $P_2+M_2$ and $M_1$ are disjoint systems, $P(P_2+_2,l,M_1,j)$ 
 is positive definite, namely, if calculated according  
 to Eq.(\ref{u15}), it is just the summand  
 on the right hand side of Eq.(\ref{u17}).

Thus we see that the initial internal state of $P_1$ determines
the outcome of the first measurement in the usual 
probabilistic sense.  One may show quite similarly that
the initial internal state of $P_2$ determines
the outcome of the second measurement in the same way.
In this sense the internal states of $P_1$ and $P_2$
play the role of local hidden variables. 
On the other hand, to hidden variables some 
supernatural features are attributed (e.g. that they cannot be measured 
in any way) and they are assumed to  
be comparable with the results of both measurements. 
None of these properties applies to the internal states of $P_1$ and $P_2$. 
As for the non-comparability we mean that the initial  
internal state of $P_1$ and $P_2$ is not comparable with 
both $|\psi{M_1}>$ and $|\psi_{M_2}>$ (while they can be compared with 
one of them). This means that 
 in our theory
there is no way to define the joint probability
$P(P_1,l_1,P_2,l_2,(0);M_1,j,M_2,k,(t))$, i.e., the probability that initially
$|\psi_{P_1}>=|\phi_{P_1,l_1}>$ and $|\psi_{P_2}>=|\phi_{P_2,l_2}>$
{\em and} finally $|\psi_{M_1}>=|m^{(1)}_j>$ 
and $|\psi_{M_2}>=|m^{(2)}_k>$. Intuitively we would write
\begin{eqnarray}
P(P_1,l_1,P_2,l_2,(0);M_1,j,M_2,k,(t))\mbox{\hspace{4cm}}\nonumber\\
=|c_{l_1}|^2\delta_{l_1,l_2}|<\xi(P_1,j)|\phi_{P_1,l_1}>|^2
|<\xi(P_2,k)|\phi_{P_2,l_2}>|^2\;,\label{u18}
\end{eqnarray}
as $|c_{l_1}|^2\delta_{l_1,l_2}$ is
the joint probability that $|\psi_{P_1}>=|\phi_{P_1,l}>$ 
and $|\psi_{P_2}>=|\phi_{P_2,l}>$, and $|<\xi(P_i,j)|\phi_{P_i,l_i}>|^2$ is the conditional
probability that one gets the $j$-th result in the $i-th$
measurement if initially $|\psi_{P_i}>=|\phi_{P_i,l_i}>$ ($i=1,2$).
Certainly the existence of such a joint probability would immediately imply the
validity of Bell's inequality, thus it is absolutely important 
to understand why this probability does not exist.
 
As our postulates provide us with equal time joint probabilities, 
we should try to express $P(P_1,l_1,P_2,l_2,(0);M_1,j,M_2,k,(t))$ 
with them. As above, we may note that the initial internal state 
of $P_1$ is uniquely related to the final internal state of $P_1+M_1$, 
 the initial internal state 
of $P_2$ is uniquely related to the final internal state of $P_2+M_2$, 
and  the final internal state of $P_1+M_1$  
is uniquely related to the final internal state of $P_2+M_2$. 
Therefore, if $P(P_1,l_1,P_2,l_2,(0);M_1,j,M_2,k,(t))$ exists, 
it coincides with $P(P_1+M_1,l_1,M_1,j,M_2,k)\delta_{l_1,l_2}$. 
But the systems $P_1+M_2$, $M_1$ and $M_2$ are not disjoint, 
neither are their complementary systems. As a result, 
if one tries to apply Eq.(\ref{u5}), one obtains 
\begin{eqnarray} 
P(P_1+M_1,l_1,M_1,j,M_2,k) 
=\sum_{j'}<\xi(P_1,j')|\phi_{P_1,l_1}><\phi_{P_1,l_1}|\xi(P_1,j)>\nonumber\\ 
\times
\left(\sum_{l'}
  c_{l'}<\xi(P_1,j)|\phi_{P_1,l'}><\xi(P_2,k)|\phi_{P_2,l'}>\right)\nonumber \\
\times
\left(\sum_{l''} c^*_{l''}<\xi(P_1,j')|\phi_{P_1,l''}>^*<\xi(P_2,k)|\phi_{P_2,l''}>^*\right) 
\label{wrongcorr} 
\end{eqnarray}  
This expression fails to be real and positive when Bell's 
inequality is violated. This can be seen because summing Eq.(\ref{wrongcorr})  
over $l_1$ one gets the correct joint probability 
\begin{eqnarray}
P(M_1,j,M_2,k)
=\left|\sum_l c_l<\xi(P_1,j)|\phi_{P_1,l}><\xi(P_2,k)|\phi_{P_2,l}>\right|^2
\quad.\label{u19}
\end{eqnarray} 
which can also be obtained directly from Eq.(\ref{u5}).
This is the usual quantum mechanical expression 
which violates Bell's inequality and whose correctness is experimentally
proven.  
 
The other side of the non-comparability is that a nondisturbing measurement 
of $|\psi_{P_1+M_1}>$ inevitably disturbs $|\psi_{P_1+M_1+P_2+M_2}>$, 
and thus $P(M_1,j,M_2,k)$, too. Therefore, if one measures $|\psi_{P_1+M_1}>$ , 
$|\psi_{M_1}$ and $|\psi_{M_2}$ (in order to give an operational definition 
for $P(P_1+M_1,l_1,M_1,j,M_2,k)$) $P(M_1,j,M_2,k)$ will be changed.  
Indeed, after having measured  
$|\psi_{P_1+M_1}>$  
\footnote{ 
This is equivalent by recording the initial internal state 
of $P_1$.} by a further measuring device $M_3$ we get for the 
internal state  of the whole system 
 \begin{eqnarray} 
\sum_l c_l\left( \sum_j <\xi(P_1,j)|\phi_{P_1,l}>|\xi(P_1,j)>|m^{(1)}_j>\right) 
\mbox{\hspace{2.5cm}}\nonumber\\ 
\times\left( \sum_k <\xi(P_2,k)|\phi_{P_2,l}>|\xi(P_2,k)>|m^{(2)}_k> 
\right)|m^{(3)}_l>\;.\label{u21} 
\end{eqnarray}  
As the systems $M_1,\;M_2,\;M_3$ are disjointed, 
we may apply {\bf Postulate 9} for $n=3$ and we get for  
$P(M_3,l_1,M_1,j,M_2,k)$ the positive definite expression 
  \begin{eqnarray} 
P(M_3,l_1,M_1,j,M_2,k)=|c_{l_1}|^2 |<\xi(P_1,j)|\phi_{P_1,l_1}>|^2 
|<\xi(P_2,k)|\phi_{P_2,l_1}>|^2\;. 
\end{eqnarray} 
This readily implies Bell's inequality. Indeed, if {\bf Postulate 9} 
is applied for Eq.(\ref{u21}) one gets
\begin{eqnarray}
P(M_1,j,M_2,k)=
\sum_l |c_l|^2 |<\xi(P_1,j)|\phi_{P_1,l}>|^2
|<\xi(P_2,k)|\phi_{P_2,l}>|^2\label{u20}
\end{eqnarray}
 which satisfies Bell's inequality and differs from
Eq.(\ref{u19}). This is due to the extra measurement which is 
equivalent with a measurement of the "hidden variable".  
 
Summarizing, we have seen that the initial internal state 
of $P_1$ ($P_2$) determines the first (second) measurement 
process, therefore, these states 'carry' the initial correlations 
and 'transfer' them to the measuring devices.  
As the measurement processes do not influence each other,  
the observed correlations may stem only from the 'common past' 
of the particles. 
On the other hand, any attempt to 
compare the initial internal states of $P_1$ and $P_2$ with 
the results of both measurements changes the correlations, 
thus a joint probability for the simultaneous existence of these states 
cannot be defined. This means that the reason for the 
 violation of Bell's inequality is that the usual derivations 
 always assume that the states (or "stable properties", "hidden variables" etc.) 
 which carry the initial correlations can be freely compared with the results 
 of the measurements. This comparability is usually  
 thought to be a consequence of realism. 
 According to the present theory, the above assumption 
 goes beyond the requirements of realism and proves to be wrong, 
 because each of the states $|\psi_{P_1+M_1}>$,  
$|\psi_{P_2+M_2}>$, $|\psi_{M_1}>$ and $|\psi_{M_2}>$ exists individually, 
but they cannot be compared without changing the correlations. 
 
\section{Summary and conclusion}  
 
A new approach to quantum mechanics was presented.  
The measurement postulates were abandoned and replaced by 
different ones to get a self-consistent, universally valid 
quantum mechanics. The present approach was based on the idea  
of quantum reference systems and the relational nature 
of quantum states, and, from the mathematical point of view 
the Schmidt representation (eigenstates of density matrices)  
was utilized. It was briefly shown how this idea 
solved the objectification problem, then the concept 
of non-comparability was discussed. Finally, it was shown 
that locality was maintained in a Bell-type experiment  
and the violation of Bell's inequality was rendered possible 
by non-comparability of existing states.   
   
According to the present approach the principle of locality 
is valid in quantum mechanics, and it is the concept of realism 
that should be modified. This modification amounts to  
surrendering the overall comparability of existing things (states) 
which are defined with respect to different (quantum) reference 
systems.  
 
Finally, let us mention that the present approach is self-contained, 
i.e., it does not rely upon the concept of {\em a priori} 
classical objects. This means that (if correct) it should give account 
of classical behavior. This is a great 
challenge but it has not been discussed here.   

\section{Acknowledgements}

The author is indebted to D.Dieks, N.D.Mermin, P.Mittelstaedt,
 L.E.Szab\'o and P.Vermaas for useful discussions 
 and remarks.
 This work has been supported by the Hungarian Aca\-demy of 
 Sciences
 under Grant No. OTKA T 029752 and the J\'anos Bolyai Research Fellowship.

\end{document}